\documentclass[aps,physrev,onecolumn,groupedaddress]{revtex4-2}

\usepackage{graphicx}
\usepackage{dcolumn}
\usepackage{bm}

\usepackage{color}
\usepackage{amsmath}
\usepackage{amssymb}

\graphicspath{{figures/}}
\usepackage{cleveref}

\newcommand{\lz}[1]{{\color{black}{#1}}}
\newcommand{\JP}[1]{{\color{black}  #1}}

\begin{document}


\title{\JP{A neural network architecture and training algorithm to predict viscoelastic stresses from vortical data}}

\author{Lu Zhu}
\author{Jacob Page}%
 \email{Contact author: Jacob.Page@ed.ac.uk}
\affiliation{%
 School of Mathematics and Maxwell Institute for Mathematical Sciences, University of Edinburgh, Edinburgh, EH9 3FD,UK
}%


\date{\today}

\begin{abstract}
\JP{
Numerical simulations of elastic turbulence in parallel shear flows of polymer solutions indicate that the phenomena is associated with the formation and instability of exact coherent states dominated by thin sheets of polymer stress. 
However, these ``arrowhead’’ structures are yet to be seen directly in experiments, where simultaneous velocity and polymer conformation measurements are challenging to obtain. 
Motivated by these challenges, we introduce a method for the prediction of the polymer conformation field given a time series of vorticity measurements. 
Our approach consists of two components: the first is a convolutional neural network architecture which takes vorticity fields and outputs a positive definite conformation tensor. 
The second is the adaptation of an assimilation-based training algorithm \citep{Zhu_Page2026} which does not require a pre-generated `offline’ library of reference conformation fields, but is trained only using the vorticity measurements. 
This is particularly important in viscoelastic problems, where the appropriate model and parameters to compare to the experiments may need to be determined as part of the solution. 
In training, measurements made on a time-marched network prediction are required to match the saved time series, while the output of the solver and network predictions at later times are required to be self-consistent.
We apply these ideas to two-dimensional Kolmogorov flow in a range of regimes, from simple traveling waves to a fully chaotic state. 
In all cases, our method produces robust predictions of the polymer stretch, while standard, unregularised variational assimilation is ineffective. 
In the chaotic case we show that our networks generalise to much larger domains -- without further optimisation -- than the `minimal’ units in which they were trained. 
}
\end{abstract}


\maketitle

\section{Introduction}\label{sec:intro}

The elasticity introduced with the addition of small amounts of polymer additives to a viscous Newtonian fluid can substantially influence flow dynamics across a range of dynamical regimes \citep{white2008mechanics,dubief2023elasto}. 
In inertia-dominated flows, viscoelasticity is associated with dramatic reductions in wall drag~\citep{virk1975drag,white2008mechanics,xi2019turbulent}.
This effect is particularly beneficial in industrial applications involving the long-distance transport of fluids through pipes~\citep{karami2012investigation}. 
On the other hand, in inertialess flows, elastic stresses can trigger elastic instabilities which may subsequently develop into elastic turbulence~\citep{groisman2000elastic,steinberg2021elastic}.
Such elasticity-driven chaotic motion enhances fluid mixing and can therefore improve heat transfer and mass exchange~\citep{kurzthaler2021geometric}.

\JP{
While earlier work suggested the need for base-flow curvature to trigger inertialess instabilities to seed ET \citep{Larson1990,shaqfeh1996purely}, self-sustaining inertialess chaos has been observed in two- \citep{berti2008two,lewy2025revisiting} and three-dimensional \citep{lellep2024purely} simulations of \emph{parallel} shear flows. 
In all cases cited here, the appearance of ET at zero or near-zero Reynolds numbers is associated with an instability of a finite-amplitude traveling wave solution \citep{lellep2023linear} which features large amplitude sheets of polymer stretch meeting at local maximum in the base velocity \citep[note the exceptional examples of chaos which is linked to a diffusive wall mode][]{beneitez2023,beneitez2025}. 
These `arrowhead’ structures are connected to a centre-mode linear instability recently discovered in pipes and channels \citep{garg2018viscoelastic,khalid2021b}.
The traveling waves are strongly subcritical \citep{page2020exact,morozov2022coherent,buza2022finite,zhu2026essential} and are also observed in parts of the parameter space associated with elasto-inertial turbulence \citep{dubief2013mechanism,sid2018two}, though they are believed to be unimportant to sustaining the chaotic dynamics there \citep{beneitez2024}.
}

Although some numerically observed phenomena, such as elasto-inertial turbulence, have been indirectly verified in experiments~\citep{samanta2013elasto}, \JP{the arrowhead structure closely associated with ET in parallel flows has not been seen definitively in the laboratory. 
The most promising results are perhaps those of \citep{choueiri2021experimental}, which observed the signature of a centre mode velocity field in a pipe. 
However, the defining polymer sheets that would confirm such a structure are more challenging to observe. 
This challenge is also reflected in past experimental investigations of viscoelastic instabilities, which typically focus on measurements of the flow kinematics \citep[e.g. in the investigation of ET in straight channels][]{Pan2013,Qin2017}.}
Complementary information about polymer stretching can be obtained using flow-induced birefringence \citep{Haward2012,haward2021bifurcations}: under a calibrated stress--optic relation, optical retardation and optical-axis orientation can quantify the in-plane deviatoric polymer stress. 
\JP{
However, combining this with time-resolved velocimetry in an unsteady flow is experimentally demanding. 
Moreover, even in an ideal planar flow it does not determine the isotropic part of the polymer stress or conformation tensor. 
Consequently, temporally co-incident, time-resolved measurements of the velocity and full polymer conformation tensor are not routinely available. 
Motivated by this, we explore here the utility of machine learning to infer the conformation field from vortical measurements, 
}

\JP{
Recent work has trained neural networks for this particular inverse problem, with a range of recent studies appearing in this area
}
\citep{balasubramanian2025prediction,thakur2024viscoelasticnet,mahmoudabadbozchelou2022nnpinns,otto2024machine,chen2026physicsguided}. 
\JP{For instance,} \citet{balasubramanian2025prediction} trained fully convolutional neural networks (CNN) on DNS data to infer instantaneous near-wall polymeric-stress fluctuations in viscoelastic turbulent channel flow from velocity fields or wall-based quantities, \JP{showing some ability to identify large-scale structures at low Weissenberg numbers.}
\JP{In another study with a similar configuration,} \citet{chen2026physicsguided} proposed a Spectral-Fourier U-Net CNN model that incorporates a soft penalty to promote the positive definiteness of the predicted tensor, and reported improved performance compared with conventional CNN architectures. 
\JP{A common feature to both these approaches is that the models were trained offline, requiring large amounts of reference stress fields. 
In an experiment, this data is not typically available. 
}
\JP{The generation of large amounts of offline training data is also unwise when the underlying model and parameters for a particular experiment might be unknown.}
\JP{The use of machine learning in model selection has been considered by a range of authors, using both carefully constrained neural networks to parameterise general admissible models \citep{Lennon2023} or physics informed neural networks \citep{thakur2024viscoelasticnet,mahmoudabadbozchelou2022nnpinns}.
Of particular relevance to the present study is the recent work from \citet{Alp2025}, who trained their model-parameterising neural networks `online' within a differentiable solver. 
}

\JP{
In this paper we present an algorithm to construct the polymer conformation field from observations of the vorticity field alone. 
We consider viscoelastic Kolmogorov flow in a range of regimes (relative equilibria, periodic orbits and fully chaotic) based on the phase diagram constructed by \citet{lewy2025revisiting} and motivated by a new experimental campaign from the Arratia group \citep{Achiriloaie2025}. 
Our approach consists of two components.
The first is a convolutional neural network that produces a positive definite conformation tensor given a short times series of vorticity snapshots. 
Positive definiteness is enforced in the network as a hard constraint, similar to the `Leray layer’ used in \citep{page2025super} for divergence-free velocity fields. 
The use of vorticity as an input is motivated by the ability of recent experiments \citep{Achiriloaie2025} to accurately construct this quantity from PIV data (P Arratia, private communication). 
The second component is a training algorithm for the predictive network which does not require access to offline conformation fields, but instead trains the model `online’: predicted polymeric quantities are time-marched with a differentiable solver. 
The solver is trained by requiring that the resulting \emph{vorticity} time series from the simulation matches the reference data \citep{page2025super,weyrauch2026state}, similar to the 4DVar assimilation algorithm, while insisting that the time-stepped conformation is consistent with predictions of the network later in time \citep{Zhu_Page2026}. 
The lack of requirement for reference conformation data allows for iterative adjustment of properties of the flow configuration, input observable choice and polymer model, making it ideal for use with experimental data. 
}

\JP{The remainder of this paper is structured as follows.} 
In \S\ref{sec:method}, we introduce the problem formulation, datasets, and \JP{our proposed reconstruction algorithm}. 
\JP{The performance of our machine learning algorithm is analysed in \S\ref{sec:results}, where we compare it to other state-estimation methods. We examine performance for a range of attractor types, from simple traveling waves to large-domain chaotic flow.}
Finally, the main conclusions are summarized in \S~\ref{sec:concl}.

\section{Methodology}\label{sec:method}

\subsection{Physical problem and governing equations}\label{subsec:problem}

\begin{figure}
      \centering    
      \includegraphics[width=0.3\linewidth, trim=0mm 0mm 0mm 0mm, clip]{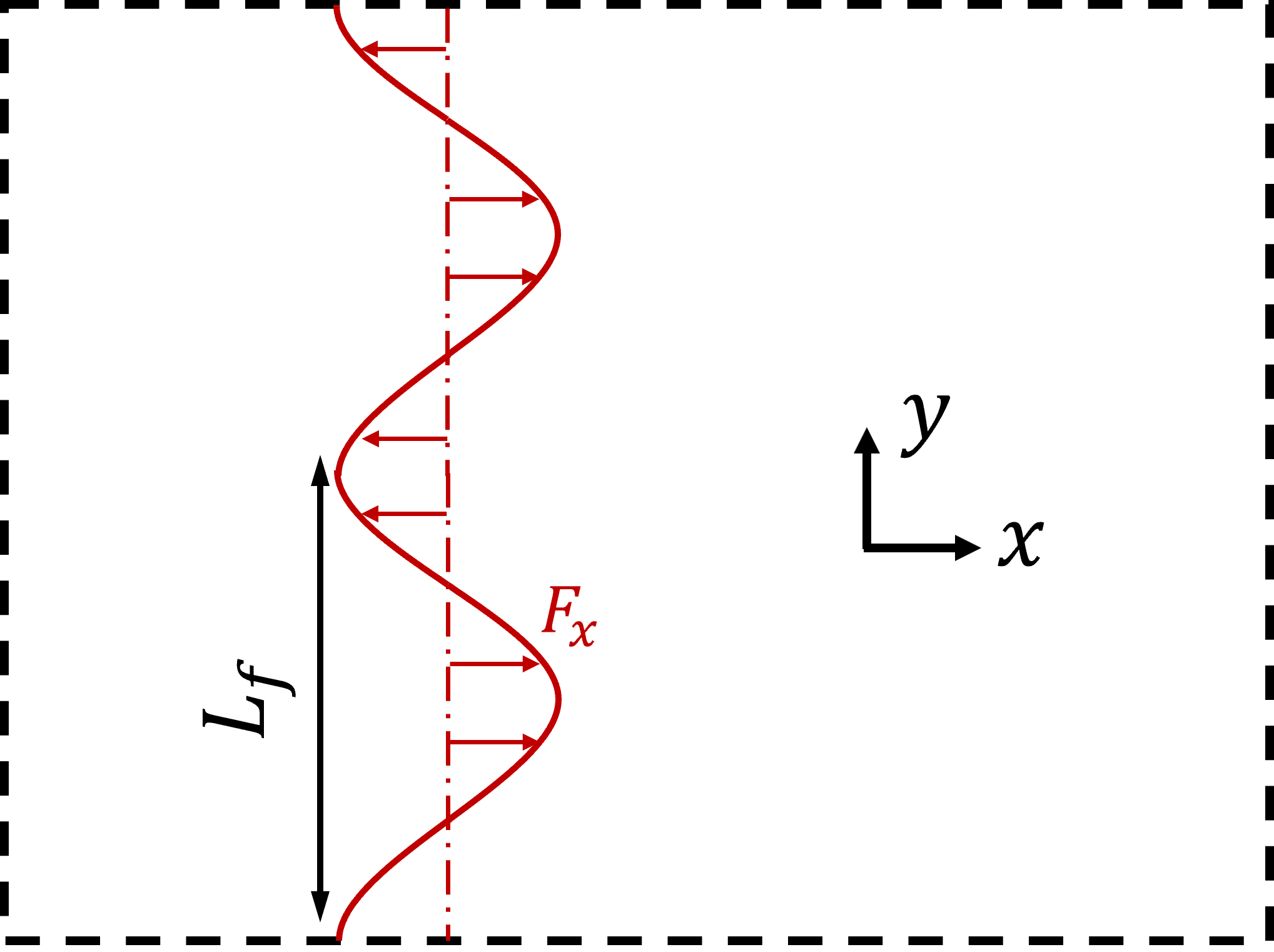}
    \caption{A schematic of Kolmogorov flow. \JP{The flow is driven by a monochromatic body force in the horizontal direction. The forcing wavelength is used as the defining lengthscale in the problem, and we consider a variety of domain sizes.}}
    \label{fig:schematic}
\end{figure}

We consider two-dimensional Oldroyd-B flow in the Kolmogorov configuration~\citep{lewy2025revisiting,zhu2026essential}, driven by a unidirectional sinusoidal body force applied in the streamwise ($x$) direction. 
The computational domain is doubly periodic. Velocities and lengths are nondimensionalized by the maximum horizontal laminar velocity $\mathcal{U}$ and the characteristic length scale is $\mathcal{L}=L_f/(2\pi)$, where $L_f$ denotes the forcing wavelength. 
A schematic of the flow geometry is shown in figure \ref{fig:schematic}.
The nondimensional governing equations for the velocity field $\bm{u}$, pressure $p$ \JP{and conformation tensor $\boldsymbol \alpha$ are}
\begin{eqnarray}
    \label{eq:ns:mass}%
    \bm{\nabla} \cdot \bm{u} = 0,
    \\
    \label{eq:ns:mom}%
    \frac{\partial \bm{u}}{\partial t} + \bm{u} \cdot \bm{\nabla} \bm{u} =
    - \bm{\nabla}p + \frac{\beta}{Re} \nabla^{2}\bm{u} +\frac{1-\beta}{Re\,Wi}\bm{\nabla}\cdot\bm{\alpha} + \frac{1}{Re}\frac{1+\epsilon\beta Wi}{1+\epsilon Wi}\cos(k_f\, y)\bm{e}_x,
    \\
    \label{eq:ns:oldb}
    \frac{ \partial\bm{\alpha}}{\partial t} + \bm{u}\cdot\bm{\nabla}\bm{\alpha} - (\bm{\nabla}\bm{u})\bm{\alpha} - \bm{\alpha}(\bm{\nabla}\bm{u})^T= -\frac{\bm{\alpha}-\bm{I}}{Wi}
    +\epsilon\nabla^2\bm{\alpha}.
\end{eqnarray}
Here, $Re\equiv\mathcal{U} \mathcal{L}/\nu$ is Reynolds number ($\nu$ is the total kinematic viscosity of the polymer solution), $Wi\equiv \lambda\mathcal{U}/\mathcal{L}$ is the Weissenberg number ($\lambda$ is polymer relaxation time), $\beta\equiv\nu_s/\nu$ is the solvent-to-total viscosity ratio and $k_f=1$ is the forcing wavenumber. 
The rightmost term in Eq.~\ref{eq:ns:oldb} is a polymer stress diffusion term~\citep{ElKareh89} which is generally small. 
In this study, we set $\epsilon=0.001$ to be consistent with recent work in the same configuration \citep{lewy2025revisiting,zhu2026essential}.

\JP{
Equations (\ref{eq:ns:mass}-\ref{eq:ns:oldb}) are solved using a modified version of the fully differentiable JAX-CFD solver \citep{kochkov2021machine,dresdner2022learning}. 
We take the curl of (\ref{eq:ns:mom}) to work with the out-of-plane vorticity, $\omega := \partial_x v - \partial_y u$; velocity is computed via solution of the Poisson problem $\nabla^2 \psi = -\omega$, with $(u,v) = (\partial_y \psi, -\partial_x \psi)$ satisfying incompressibility automatically. 
Fourier decomposition is used in both horizontal directions while time discretisation is performed using an \lz{RK2} scheme for the bi- and nonlinear terms, which semi-implicit Crank-Nicolson is used for the linear terms. 
The solver was validated by reproducing results from \citet{lewy2024revisiting} and \citet{zhuJFM2026on}. 
Importantly, the fact that the code is fully differentiable allows us to differentiate through the time forward map, $\boldsymbol \varphi_t(\mathbf q)$, with respect to initial conditions $\mathbf q$ (where $\mathbf q = (\omega, \alpha_{xx}, \alpha_{xy}, \alpha_{yy})$ is the state vector whose components consist of the vorticity and flattened conformation fields).
}

\subsection{Database}\label{subsec:database}
\JP{In recent work}, \citet{lewy2025revisiting} identified a series of elastic chaotic states in the inertialess regime.
\JP{As well as the two-dimensional elastic turbulence discovered by \citet{berti2010}, they document other simpler dynamical regimes by varying $Wi$ and the domain size. 
We select one example from each -- a traveling wave, a periodic orbit and a chaotic state -- in which we attempt to train networks to construct $\boldsymbol \alpha$ given $\omega$ given only a vorticity time series.
Details of each of the flow configurations and vorticity datasets used in training are provided in table \ref{tab:cases}.
Note that the traveling wave and periodic orbit are found at the same parameter point, where the flow exhibits bistability.
Later in the paper we will also explore the ability of our networks to estimate stresses on much larger domains, without any re-training of the model.
}

\begin{table}
    \begin{center}
    \def~{\hphantom{0}}
	{
	\begin{tabular}{l|c|cccc|c|c|c}
        Name & Regime & $\mathrm{Re}$ & $\mathrm{Wi}$ & $\beta$ & $\epsilon$ & $T\times L_x\times L_y$ & $N_t\times N_x\times N_y$  & $(w_1, w_{21}, w_{22})$ \\
        \hline
        TW & Traveling wave & 0.5 & 20 & 0.95 & 0.001 & $15\times 2\pi\times2\pi$ & $60\times 64 \times 64$  & (1,1/2,1/2)\\
        PO & Periodic orbit & 0.5 & 20 & 0.95 & 0.001 & $60\times 2\pi\times 2\pi$ & $240\times 64 \times 64$ & (1,1/2,1/2) \\
        CS & Chaotic state & 0.5 & 30 & 0.95 & 0.001 & $400\times 6\pi\times 4\pi$ & $400\times 192 \times 128$ & (1,1,1)\\
	\end{tabular}
	\caption{\JP{Parameter settings for each of the cases considered here, along with details of the training dataset}. Here, $T$, $L_x$, and $L_y$ are the duration of the dataset and the streamwise and spanwise domain lengths, respectively, while $N_t$, $N_x$, and $N_y$ denote the corresponding numbers of points in time and in the two spatial directions.  \JP{The hyperparameters for the TraCTra loss function (\ref{eq:loss:z26}) are reported in the final column.}}\label{tab:cases}
	}
	\end{center}
\end{table}

\subsection{Trajectory-consistent network training}\label{subsec:PReTra}
\JP{
We consider the problem of estimating $\boldsymbol \alpha$ from vorticity measurements, given a forward model (i.e. assuming parameters are known). 
All methods examined here do not require generation or use of `offline' conformation data. 
This is key in viscoelastic flows, since (i) it allows for change of the model in the simulation without regenerating any data and (ii) it is known that multiple attractors can coexist at the same parameter point \citep{beneitez2024,lewy2025revisiting}.
}
\JP{
To do this we adapt the recently developed `trajectory consistent network training' algorithm \citep[TraCTra, see][]{Zhu_Page2026}. 
This algorithm is itself a modification of a 4DVar-inspired network training algorithm \citep{page2025super}, which we refer to here as `P25'. 
We will compare our TraCTra approach to P25 and also directly to 4DVar; we detail each of these methods here. 
}

For the problem considered in this study, the standard 4DVar approach \citep{foures2014data,wang2021state} seeks an optimal initial state 
$\tilde{\mathbf q}_0=(\tilde{\omega}_0,\tilde{\alpha}_{xx,0},\tilde{\alpha}_{xy,0},\tilde{\alpha}_{yy,0})\in \mathbb{R}^{N_x\times N_y\times 4}$ ($\tilde{\bullet}$ denotes the predicted state), 
such that its forward-time trajectory, $\boldsymbol \varphi_t(\tilde{\mathbf q_0})$, obtained through the time-forward mapping best matches the corresponding vorticity observations 
$\bm{\omega}^*\in \mathbb{R}^{N_T\times N_x/M\times N_y/M}$ 
\JP{at a set of discrete times which form} the assimilation window $t_k\in\{0,\Delta t_C,...,(N_T-1)\Delta t_C \}$. 
Here, $\Delta t_C$ is the time interval between two successive observation times and $N_T$ denotes the number of temporal snapshots within the assimilation window $T_e \equiv (N_T - 1)\Delta t_C$.
\JP{While here all observations are obtained from precursor simulations, the algorithm has been designed with experimental measurements in mind.
With $M>1$ we can also consider coarse-graining observations in space, and while this is not the primary focus of our study we do include some of these results below.} 

\JP{The assimilated in initial state is found by minimization of the loss} 
\begin{equation}
    \mathcal{L}_\mathrm{4DVar}(\mathbf q_0)=\frac{1}{N_T}\sum^{N_T-1}_{k=0}\|\omega^*_{t_k}-\mathcal{M} \circ \boldsymbol \varphi_{t_k} (\mathbf q_0) \|^2,
    \label{eq:loss:4dvar}
\end{equation}
with $\tilde{\mathbf q}_0 := \arg \min \mathcal{L}(\mathbf q_0)$, and where the observation \JP{or measurement} operator 
$\mathcal{M}:\mathbb{R}^{N_x\times N_y\times C}\rightarrow 
\mathbb{R}^{N_x/M\times N_y/M}$ 
maps the full high-resolution state to the corresponding (\JP{possibly} coarse-grained) vorticity \JP{for comparison to the measurements $\{\omega_{t_k}^*\}$}. 
\JP{With our differentiable solver, gradients $\boldsymbol \nabla_{\mathbf q}\mathcal L$ can be computed using standard JAX functionality. This is also exploited in the network training described below.}

Motivated by 4DVar, \JP{the `P25' algorithm is} a network-based training method for reconstructing high-resolution initial conditions, \JP{which seeks to combine the benefits of time marching for trajectory comparison in 4DVar with the sampling benefits of network training exposed to a wide variety of dynamics. This method was demonstrated in super-resolution problems in two- and three-dimensional turbulence \citep{page2025super,weyrauch2026state}.} 
In analogy with P25, we use a convolutional neural network (CNN), $\mathcal{N}$, to map vorticity observations to an estimate of the full initial state $\tilde{\mathbf q}_0 = \mathcal N(\Omega_0^*)$.
\JP{We obtain the network estimate using a small time series as input to the network, rather than a single snapshot, $\Omega_0^*:=\{\omega^*_{t_{-2}}, \omega^*_{t_{-1}}, \omega^*_{t_{0}}, \omega^*_{t_{1}}, \omega^*_{t_{2}}\}$, which are passed as a single five-channel image (further details below).}
The predicted state is evolved using the differentiable solver, and measurements on this trajectory are compared against the reference vorticity data. 
This leads to the loss function
\begin{equation}
    \mathcal{L}_\mathrm{P25}(\Theta)=\frac{1}{N_TN_S}\sum^{N_S}_{j=1}\sum^{N_T-1}_{k=0}\|\omega^{*,j}_{t_k}-\mathcal{M} \circ \boldsymbol \varphi_{t_k}\circ \mathcal{N}(\Omega^{*,j}_0)\|^2,
    \label{eq:loss:p25}
\end{equation}
where $\Theta$ identifies the weights of the neural network and $N_S$ denotes the number of observations trajectories available for training. 
By exploiting the ability of a sufficiently expressive network to infer subgrid-scale structures, the P25 approach can effectively predict physically plausible states without requiring a complete high-resolution training dataset~\citep{weyrauch2026state}. 
Nevertheless, this formulation can lose robustness in ill-constrained reconstruction problems, where sparse observations do not provide sufficient information to ensure dynamical consistency between the model prediction and the measurements. 

\JP{In recent work, \citet{Zhu_Page2026} presented a new assimilation-based training algorithm, `Trajectory Consistent Network Training' (TraCTra), which} not only matches the forward-evolved trajectory to the observations, but also enforces consistency between the forward-evolved trajectory and the network prediction. 
In this way, additional physical information from the governing dynamics is incorporated into the training process, \JP{enabling network training in more challenging state estimation tasks \citep[for example][estimated three dimensional velocity fields from density shadowgraph images]{Zhu_Page2026}}. 
\JP{We adapt these ideas to the viscoelastic problem here, with a loss function}
\begin{equation}
    \mathcal{L}(\Theta)=\frac{1}{N_TN_S}\sum^{N_S}_{j=1}\sum^{N_T-1}_{k=0} \left(
    w_1 \underbrace{\|\omega^{*,j}_{t_k}-\mathcal{M} \circ\boldsymbol \varphi_{t_k}\circ\mathcal{N}(\Omega^{*,j}_0)\|^2}_{\text{assimilation}} + w_2 \underbrace{\| \mathcal{N}(\Omega^{*,j}_{t_k}) -\boldsymbol \varphi_{t_k}\circ\mathcal{N}(\Omega^{*,j}_0)\|^2}_{\text{consistency}}
    \right),
    \label{eq:loss:z26}
\end{equation}
where $w_1$ and $w_2$ are weights of the observation/measurement matching term (i.e. the 4DVar or P25 component) and \JP{a new term which compares time-marched \emph{full states} to network predictions of full states at later times}.
\JP{Note that comparison of the second term is made in the high-dimensional full state space, but all predictions are generated online in the loss from measurements.}

\JP{In a} viscoelastic flow, \JP{typical amplitudes} of the conformation tensor and the velocity/vorticity differ greatly, so that a loss function like (\ref{eq:loss:z26}) may significantly bias reconstruction of the former.
We \JP{therefore} adopt a more balanced relative loss defined as
\begin{equation}
\begin{aligned}
\mathcal{L}_\mathrm{Z26}(\Theta) = \frac{1}{N_TN_S}\sum^{N_S}_{j=1}\sum^{N_T-1}_{k=0}\Bigg[\;
& w_1\,
\frac{\|\omega^{*,j}_{t_k}-\mathcal{M}\circ\boldsymbol\varphi_{t_k}\circ\mathcal{N}(\Omega^{*,j}_0)\|^2}
     {\|\omega^{*,j}_{t_k}\|^2} \\[4pt]
&+ w_{21}\,
\frac{\|\mathcal{M}_\omega\circ\mathcal{N}(\Omega^{*,j}_{t_k})-\mathcal{M}_\omega\circ\boldsymbol\varphi_{t_k}\circ\mathcal{N}(\Omega^{*,j}_0)\|^2}
     {\|\mathcal{M}_\omega\circ\boldsymbol\varphi_{t_k}\circ\mathcal{N}(\Omega^{*,j}_0)\|^2} \\[4pt]
&+ w_{22}\,
\frac{\|\mathcal{M}_\alpha\circ\mathcal{N}(\Omega^{*,j}_{t_k})-\mathcal{M}_\alpha\circ\boldsymbol\varphi_{t_k}\circ\mathcal{N}(\Omega^{*,j}_0)\|^2}
     {\|\mathcal{M}_\alpha\circ\boldsymbol\varphi_{t_k}\circ\mathcal{N}(\Omega^{*,j}_0)\|^2}
\;\Bigg].
\end{aligned}
\label{eq:loss:z26}
\end{equation}
where $\mathcal{M}_\omega$ and $\mathcal{M}_\alpha$ denote projection operators that extract the (full resolution) vorticity and conformation-tensor components \JP{as vectors (all norms are standard 2-norms on vectors)}, respectively.
\JP{The hyperparameters $(w_1, w_{21}, w_{22})$ are reported in table \ref{tab:cases}.}
We refer to the output of models trained with (\ref{eq:loss:z26}) as `Z26' for brevity.

All optimization procedures -- \JP{both the variational assimilation and the network training --} are performed using the Adam optimizer~\citep{kingma2015adam}, with a learning rate of $10^{-2}$ for 4DVar and $10^{-4}$ for P25 and Z26. 
The P25 and 4DVar optimizations are initialized randomly, whereas Z26 is initialized from a pre-trained P25 model to improve training efficiency. 
The network architecture is discussed in the following section.

\subsection{Network architecture}\label{sec:network}

\begin{figure}
      \centering    
    \includegraphics[width=0.97\linewidth, trim=0mm 0mm 0mm 0mm, clip]{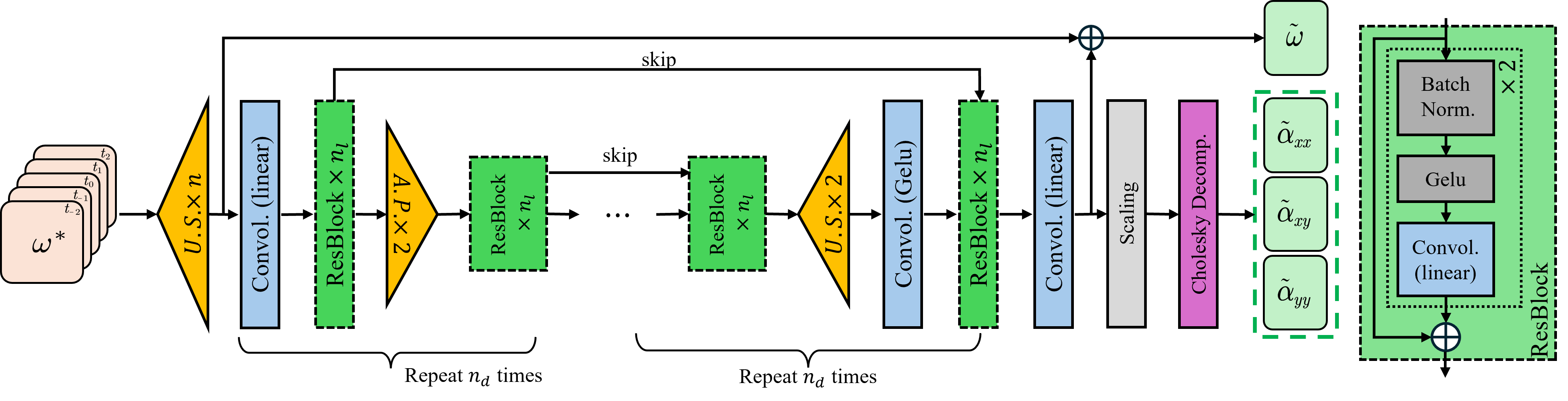}
    \caption{Network architecture of the ResUnet \JP{used in this study}. \JP{The `ResBlock' component is shown in detail on the right of the figure. Here `U.S.' indicates upsampling (necessary if coarse graining has been applied to the vorticity measurements), while `A.P.' indicates average pooling. The Cholesky layer ensures a postive definite output and is described in detail in the text.}}
    \label{fig:network}
\end{figure}

\JP{We have designed a residual-Unet network \citep[ResUNet, see][]{zhang2018road}, built around a convolutional `core', for polymer stretch reconstruction.}
The architecture of the network is illustrated in figure \ref{fig:network}. 
As described above, we use five consecutive vorticity snapshots as the `channels' in a 5-channel input image, from a stencil $t_{-2}$, $t_{-1}$, $t_0$, $t_1$, and $t_2$ around the target time $t_0$. 
The network then attempts to reconstruct the vorticity (if downsampled) and conformation tensor at $t_0$. 
\JP{In cases with $M>1$ (i.e. coarse grained in space)} input fields are first linearly interpolated to the target resolution and then mapped to $n_f$ feature channels through a convolutional layer \JP{with periodic padding to ensure satisfaction of the physical boundary conditions}.
The ResUNet employs a U-Net-type architecture~\citep{ronneberger2015u} with residual convolutional blocks. 
At all levels except the coarsest bottleneck, the feature maps are downsampled by average pooling with a factor of two, accompanied by a doubling of the channel dimension. 
The bottleneck consists of residual blocks at the lowest spatial resolution.
At each decoding level, the feature maps are upsampled by a factor of two, passed through a periodic convolutional mixing layer, and concatenated with the corresponding encoder features via a skip connection. 
The merged features are then refined by $n_l$ residual blocks.

Finally, two linear convolutional output heads map the decoded features to a residual vorticity field and three latent `conformation' fields respectively. 
\JP{The residual vorticity is added to the upsampled field just downstream of the network input (see arrows in figure \ref{fig:network}).}
\JP{The three latent `conformation' fields, $\{a(\mathbf x), b(\mathbf x), c(\mathbf x)\}$ are converted to a valid, positive definite conformation tensor field in the final `Cholesky layer' (purple box in figure \ref{fig:network}).}
\JP{Here conformation tensor is constructed pointwise as} \citep[see][]{golub2013matrix}
\begin{equation}
    \alpha_{xx}=\mathrm{softplus}(a)^2+b^2,\quad \alpha_{xy}=b\cdot\mathrm{softplus}(c),\quad \alpha_{yy}=\mathrm{softplus}(c)^2,
    \label{eq:chol}
\end{equation}
\JP{where $\mathrm{softplus}(z) := \log(1 + \exp(z))$.}
This parametrization substantially improves the convergence of the training procedure and reduces the risk of numerical failure in the forward-time solver caused by violations of the positive definiteness of the conformation tensor. 
Similarly, in 4DVar, instead of directly optimising the state vector $\tilde{\mathbf q}_0$, we instead seek a latent vector $\tilde{\bm{\phi}}_0=(\tilde{\omega}_0,\tilde{a}_0,\tilde{b}_0,\tilde{c}_0)$ from which the state vector is reconstructed through the Cholesky parametrization in \cref{eq:chol}.

In all cases considered in this study, we use $n_l=2$ residual blocks at each resolution level. The network starts with $n_f = 16$ filters, with the number of filters doubled after each encoder downsampling operation and halved correspondingly during decoding. All convolutional layers employ $(3,3)$ kernels with a stride of one. The number of resolution levels is set to $n_d=4$ for the CS cases and $n_d=3$ for all other cases (total parameters $\sim 440$k and $\sim 170$k respectively). 
The assimilation window is chosen as $T_e=5$ for the simpler dynamical states -- the traveling wave and periodic orbit -- and $T_e=15$ for the chaotic cases (TW, PO and CS in table \ref{tab:cases}), while the separation times between snapshots are $\Delta t_{TW} = 0,25$, $\Delta t_{PO} = 0.25$ and $\Delta t_{CS} = 1$.
The batch size in all cases is 11. 
\JP{When training the networks we allow assimilation windows to overlap} hence the number of subtrajectories used for training is $N_S = 37$, $N_S = 217$, and $N_S = 382$ for the traveling wave (TW), periodic orbit (PO), and chaotic states (CS) respectively.

\section{Results}\label{sec:results}
In this section we evaluate the three reconstruction methods described above: the standard 4DVar assimilation (equation \ref{eq:loss:4dvar}), neural networks trained with the `P25' algorithm (equation \ref{eq:loss:p25}), and the trajectory consistent, or `Z26' approach (equation \ref{eq:loss:z26}).
We examine their performance for the three representative dynamical states in table \ref{tab:cases}).
All three states are dominated by arrowhead structures in the polymer stretch as described in \citep{lewy2024revisiting,zhu2026essential}.
\JP{In the first instance, these are each treated as single-trajectory inverse problems, and we report results directly on the `training' data (as would be done for a PINN, for example \citep{raissi2019physics}). 
However, for or the chaotic case we examine the ability of the trained model to generalise to unseen trajectories and to changes in domain size, where even our model trained on a very limited dataset performs surprisingly effectively.}

\subsection{Traveling wave}\label{sec:tw}
The viscoelastic traveling-wave solution consists of a fixed arrowhead structure that propagates in the streamwise direction. 
The state is characterized by two symmetric polymer-stretch sheets, strongly tilted with respect to the flow direction, which are connected by an arch-like head. 
This structure is primarily manifested in the conformation field, commonly visualized through the trace of the conformation tensor, $\mathrm{tr}\,\bm{\alpha}$. 
Because the conformation field is particularly difficult to measure experimentally, direct observation of this structure remains limited despite extensive numerical observations. 

\begin{figure}
      \centering    
    \includegraphics[width=0.7\linewidth, trim=0mm 0mm 0mm 0mm, clip]{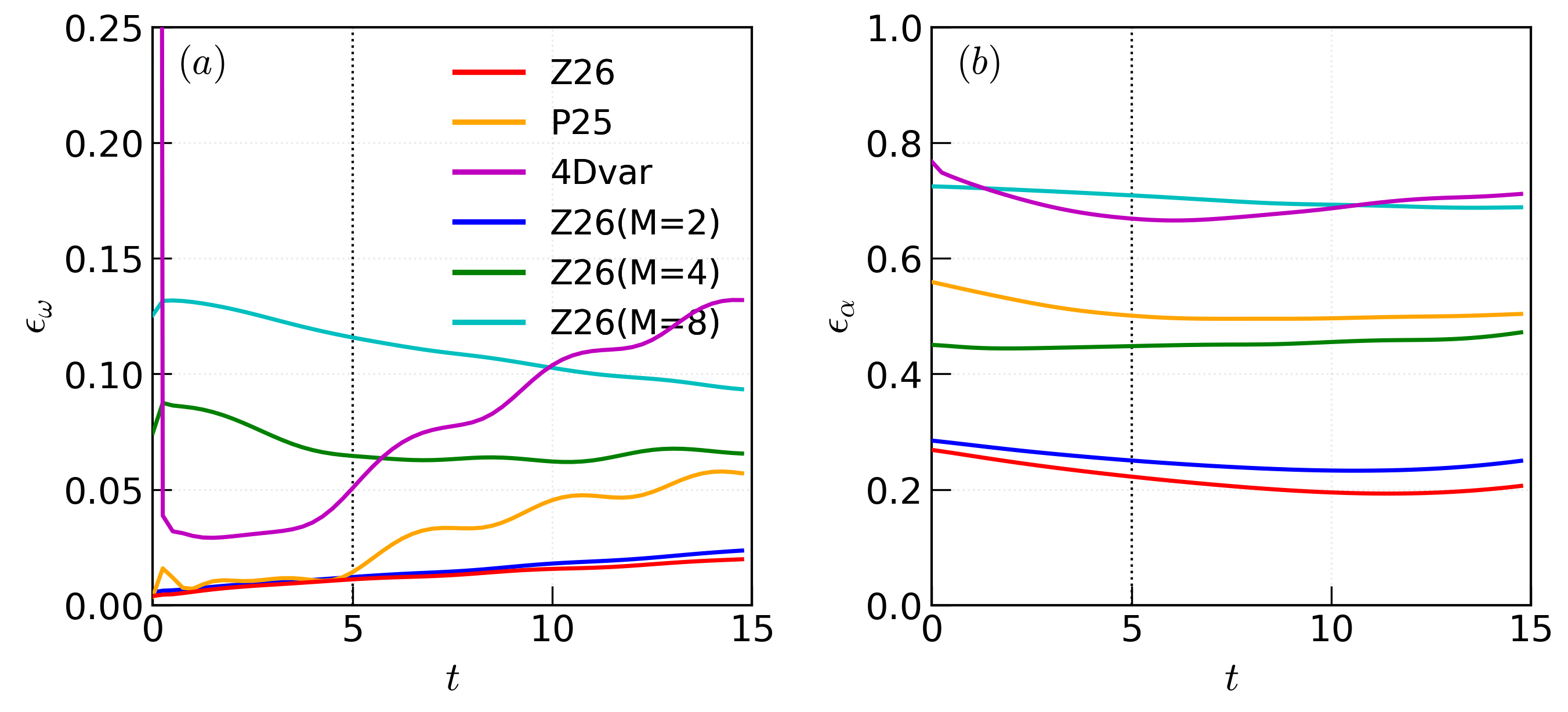}
    \caption{Temporal evolution of the relative reconstruction errors for (a) the vorticity, $\epsilon_\omega$, and (b) conformation tensor, $\epsilon_\alpha$, in the traveling-wave (TW) case. 
    The dotted line marks the assimilation time $T_e$ (\JP{which is also the time used in network training}).
    \JP{Note the figure also includes the results of the Z26 algorithm applied to downsampled vorticity fields (factors $M=2,4,8$); all other results have access to vorticity field at full spatial resolution. }
    }
    \label{fig:error_tw}
\end{figure}
In figure \ref{fig:error_tw}, we show the time evolution of the relative error of the vorticity and conformation tensor between the DNS data and the rolled out trajectory \JP{$(\hat{\omega}(t), \hat{\boldsymbol \alpha}(t)) = \boldsymbol \varphi_{t} (\mathcal N(\Omega_0^*))$,}  
\begin{eqnarray}
    \label{eq:err:vor}
    \epsilon_\omega(t)&=&\frac{\|\omega(t)-\hat{\omega}(t)\|}{\|\omega(t)\|},\\
    \label{eq:err:c}
    \epsilon_{\bm{\alpha}}(t)&=&\frac{1}{3}\sum_{c}\frac{\|\alpha_{c}(t)-\hat{\alpha}_{c}(t)\|}{\|\alpha_{c}(t)\|},
\end{eqnarray}
where $c\in\{xx, xy, yy\}$ identifies individual components of the conformation tensor $\bm{\alpha}$ \JP{and the norms are just two norms on flattened vector representations of individual conformation components}.
\JP{
These results are obtained by first evaluating the network on a single 5-timestep-stencil of the vorticity field before time marching with the solver: at runtime no further optimisation is performed. 
This is in contrast to the 4DVar results (purple lines in figure \ref{fig:error_tw}) in which an optimisation is performed over the first $t=T_e$ time units (see vertical line in the figure). 
}

\JP{
For all cases, vorticity errors are low, $\epsilon_{\omega} \leq 0.15$, over the duration of the time window. 
The exception is the 4DVar result, which starts at a very large relative error, $\epsilon_{\omega} \sim 2.9$, which rapidly decays over the assimilation window. 
These small errors in vorticity are perhaps expected since in all cases this is an input to the problem (though the optimiser can still adjust the field as it attempts to match the time evolution). 
For the Z26 algorithm we observe the smallest value of $\epsilon_{\omega}(t)$, which remains low over the entire observation window. 
Even under significant coarse-graining in space, the Z26 algorithm is competitive with 4DVar.
}

\JP{
Rising errors in the vorticity field can be attributed to errors in the conformation field. 
Note that at the low values of $Re=0.5$ under consideration, errors in the polymer stress, and hence the forcing in the momentum equation, can rapidly drive errors in the vorticity field.
}
Consistent with this interpretation, our DNS tests (not shown) indicate that even when the initial velocity/vorticity field is set to zero, the flow can be rapidly regenerated from the conformation-tensor field alone. In contrast, a vorticity/velocity field initialized with zero conformation rapidly decays toward the laminar state.

\JP{
For the conformation reconstruction, the time evolution of $\epsilon_{\alpha}(t)$ reported in figure \ref{fig:error_tw} also show the Z26 algorithm is clearly the most effective, with relative errors of around $25\%$ over the observation window.
In fact, the Z26 approach is more accurate in its conformation reconstruction than other methods even when spatial coarse-graining of up to $M=4$ is applied. 
The `trajectory consistency' terms in the loss (\ref{eq:loss:z26}) have a dramatic impact in improving the reconstruction over the measurement-assimilation-only approach of the P25 training and the 4DVar algorithm.
Note that the performance of the latter method is particularly poor ($\epsilon_{\alpha}\sim 0.75$) despite the relatively simple dynamics -- though it is worth bearing in mind that there are three attractors at this parameter point (the TW, PO and the laminar state) and there are no guarantees the assimilated state saturates onto the TW as $t\to \infty$.
}

\begin{figure}
      \centering    
    \includegraphics[width=0.97\linewidth, trim=0mm 0mm 0mm 0mm, clip]{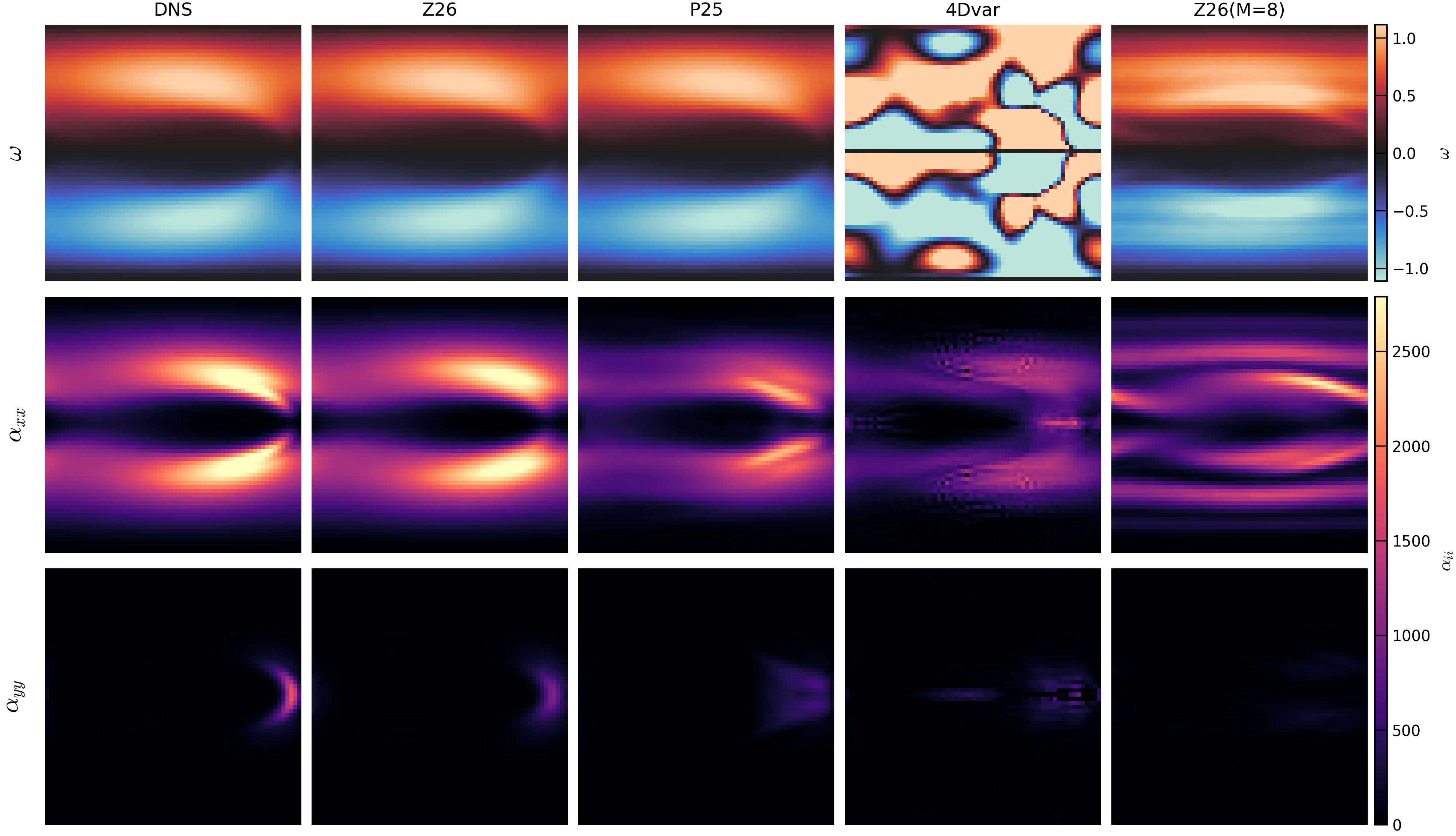}
    \caption{Comparison of network-predicted snapshots of individual flow fields with the corresponding DNS snapshots for the travelling-wave case. \JP{Rows show vorticity, streamwise-normal and vertical-normal conformation fields respectively. Columns are labeled by the corresponding state estimator.}}
    \label{fig:flowf_tw}
\end{figure}

\JP{These trends are borne out in the instantaneous fields visualised in figure} \ref{fig:flowf_tw}, where the output of the various state estimation approaches is shown alongside the DNS input. 
The Z26 prediction closely resemble the DNS across all \JP{flow variables shown}. 
The characteristic legs (see $\alpha_{xx}$) and arch-like head (contours of $\alpha_{yy}$) of the structure are clearly \JP{visible in the network outputs}. 
As the input resolution becomes coarser, the reconstruction quality deteriorates: 
\JP{For instance, in the} $M=8$ \JP{results shown in the final column there are additional sheets of $\alpha_{xx}$ predited, while the vertical conformation is no longer visible.}
\JP{
The other state estimation algorithms (P25 and 4DVar) are also unable to produce conformation fields with the qualitative features associated with the arrowhead; 
there are multiple sheets of $\alpha_{xx}$ in the outputs while the vertical conformation is substantially weaker than the DNS baseline. 
}

\begin{figure}
      \centering    
    \includegraphics[width=0.9\linewidth, trim=0mm 0mm 0mm 0mm, clip]{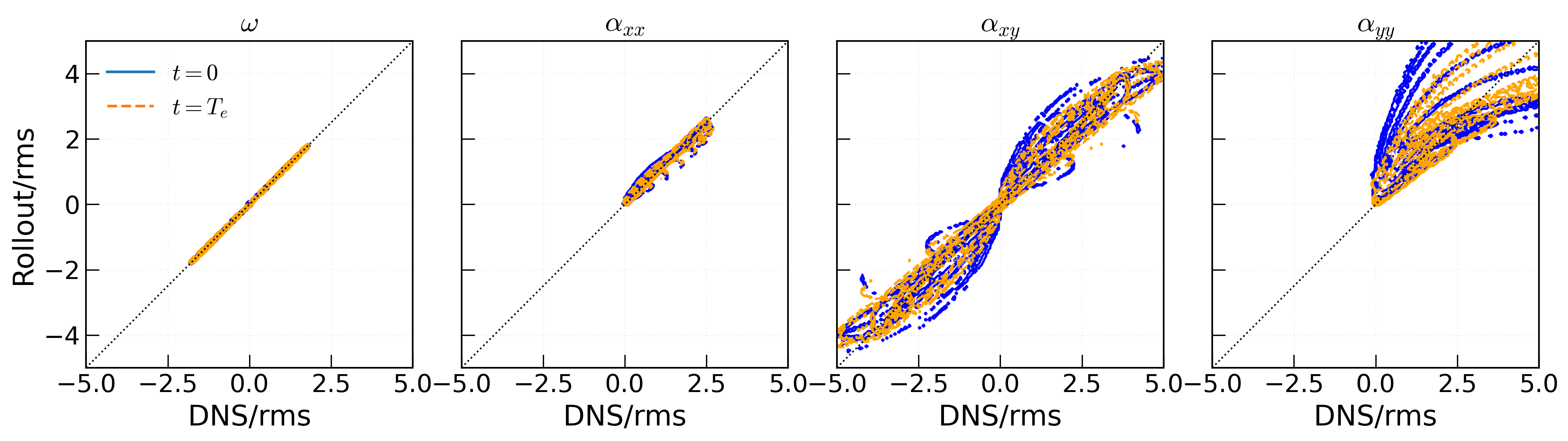}
    \caption{Joint PDFs between the DNS fields and rollout predictions for the traveling-wave case at $t=0$ and $t=T_e$. All quantities are normalized by the root-mean-square value of the corresponding DNS field. The black dashed line marks the one-to-one line \JP{on which perfect predictions would lie}.}
    \label{fig:jpdf_tw}
\end{figure}
\JP{The raw `conformation error' $\epsilon_{\alpha}$ equally weights reconstruction of all components, while in reality dominant component is the first normal value $\alpha_{xx}$.
Motivated by this, the correlation between the outputs of the Z26 approach and the DNS data is examined in  
}
figure \ref{fig:jpdf_tw}, where we present the component-wise joint probability density functions (JPDFs) between the DNS fields and the Z26 reconstructions, \JP{both at initial time $t=0$ and after time stepping to $t=T_e$}. 
As shown, both $\omega$ and $\alpha_{xx}$ are \JP{reproduced} accurately, with their contours closely aligned with the one-to-one line. 
We also note that the contours of the diagonal conformation components, $\alpha_{xx}$ and $\alpha_{yy}$, remain positive, consistent with the positive-definite constraint imposed by the Cholesky parametrization layer.
Compared with $\alpha_{xx}$, the contours of $\alpha_{xy}$ and $\alpha_{yy}$ are more broadly distributed in the direction normal to the one-to-one line, indicating lower reconstruction accuracy for these components. 
This reflect an imbalance in the component-wise contributions to the loss function during training. 
Since $\alpha_{xx}$ typically has a larger magnitude than the other conformation components, the optimizer preferentially reduces the error in $\alpha_{xx}$ when a uniform weight is applied to all conformation components. 
Potential improvement may be achieved by introducing a balanced component-wise weighted loss.

Finally, \JP{we have observed that the} Cholesky layer is not essential for maintaining the stability of the forward-time integration \JP{for the simple traveling wave state considered here (not shown)}. 
However, the issue becomes more severe for the chaotic state, where violations of the positive definiteness of the conformation tensor are more likely to destabilize the forward solver.

\subsection{Periodic orbits}\label{sec:po}

\begin{figure}
      \centering    
    \includegraphics[width=0.7\linewidth, trim=0mm 0mm 0mm 0mm, clip]{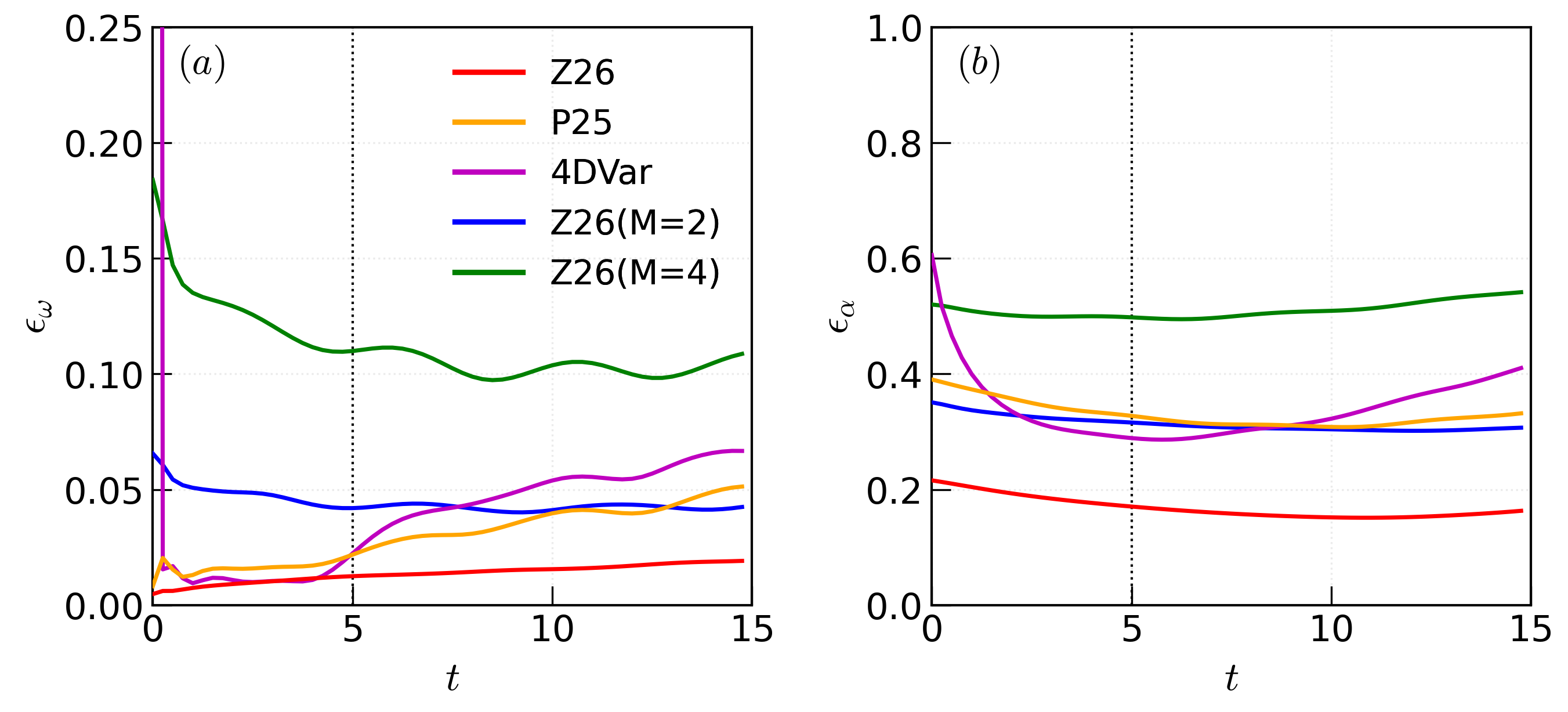}
    \caption{Temporal evolution of the relative reconstruction errors for the vorticity, $\epsilon_\omega$, and conformation tensor, $\epsilon_\alpha$, in the periodic orbit (PO) case. The dotted line marks the assimilation time $T_e$.}
    \label{fig:error_po}
\end{figure}

The periodic-orbit state (PO) is characterized by a pair of oppositely oriented arrowhead structures that propagate in opposite directions \JP{and coexists as an attractor at the same parameter point as the TW considered above}. 
\JP{We collapse onto the PO by initialising our DNS with a} perturbation consisting of spanwise-aligned stretched polymer structures -- \JP{see detail in} \citet{zhu2024early}. 

In figure \ref{fig:error_po}, we report the temporal evolution of the reconstruction errors, $\epsilon_{\omega}$ and $\epsilon_{\alpha}$, for the PO. 
Similar to \JP{the simpler} TW \JP{considered above}, the vorticity errors of both the Z26 and P25 approaches remain small, below $0.01$, during the rollout interval $0\leq t\leq 5$. 
In contrast, 4DVar exhibits an extremely large initial vorticity error at $t = 0$, \JP{which drops rapidly, similar to the behaviour in the TW case above}.

\JP{For the conformation tensor, the Z26 algorithm yields a state estimator which is again clearly superior to all other methods; $\epsilon_{\alpha} \lesssim 0.2$ consistently under time marching.
Even under coarse graining, the approach is still comparable with the other state estimation methods (4DVar and the P25 training algorithm) which have access to the full vorticity field. 
}

\begin{figure}
      \centering    
    \includegraphics[width=0.97\linewidth, trim=0mm 0mm 0mm 0mm, clip]{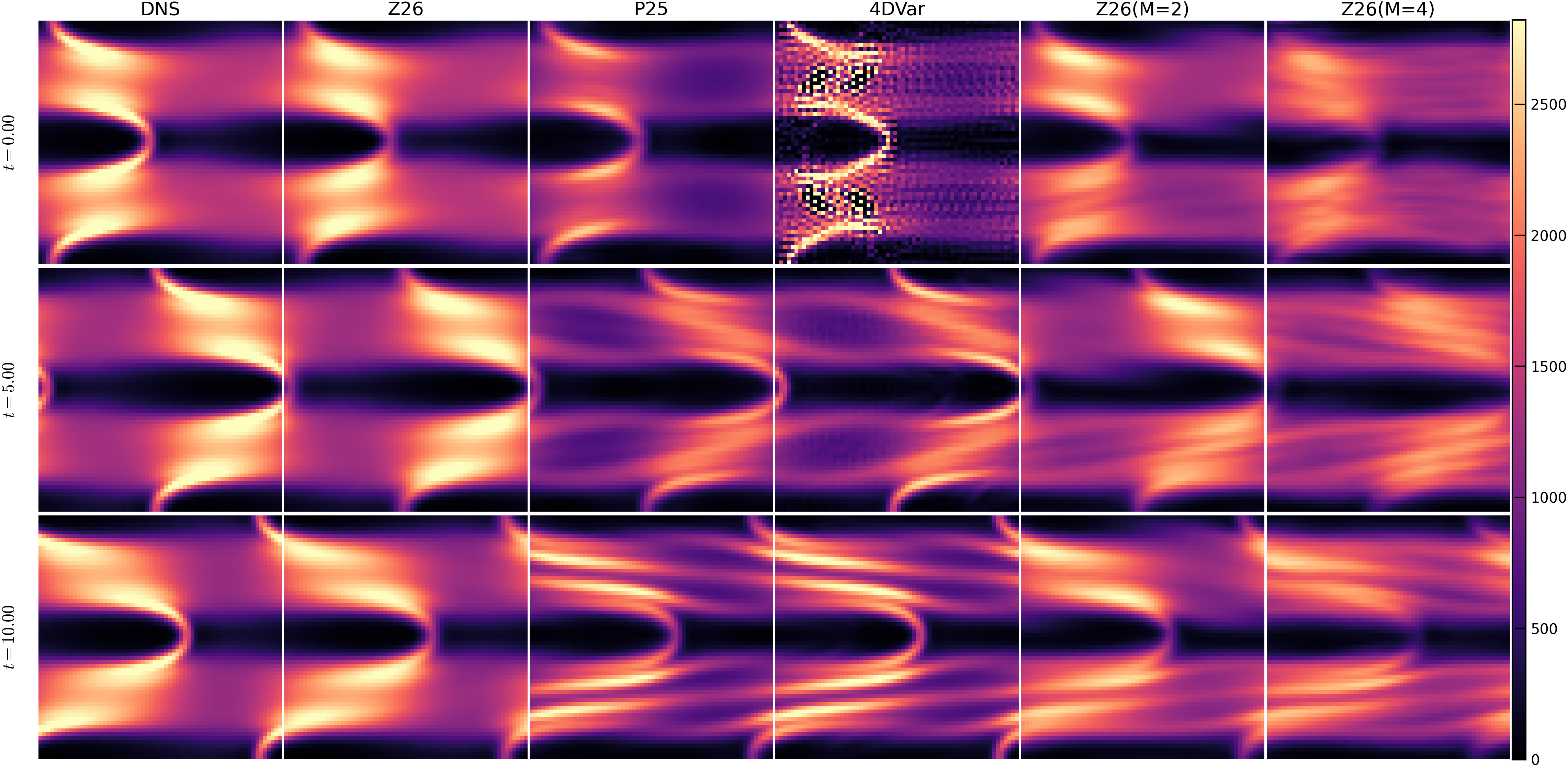}
    \caption{Temporal evolution of the trace of the conformation tensor, $\mathrm{tr}\,\bm{\alpha}$, along the rollout trajectory for the periodic orbit case (with period $T\approx 17.5$) at $t=0$, $5(T_e)$, and $10(2T_e)$.}
    \label{fig:flowf_t_po}
\end{figure}
\JP{
The low errors in the Z26 approach translate to conformation fields which retain all the key features of the reference DNS data -- see the time series of $\text{tr}\,\boldsymbol \alpha$ in figure \ref{fig:flowf_t_po}. 
We see that the neural network output cleanly produces the counter-propagating pair of arrowhead structures, which are retained under time evolution.
On the other hand, the alternative state estimation approaches struggle to match the amplitude of the trace in the true solution, and even a qualitative comparison to the DNS data degrades as the states are time marched (see e.g. the increased numbers of sheets). 
The same is also true of Z26 under coarse graining, hence access to the full vorticity field is important for the success of the approach. 
}

\begin{figure}
      \centering    
    \includegraphics[width=0.9\linewidth, trim=0mm 0mm 0mm 0mm, clip]{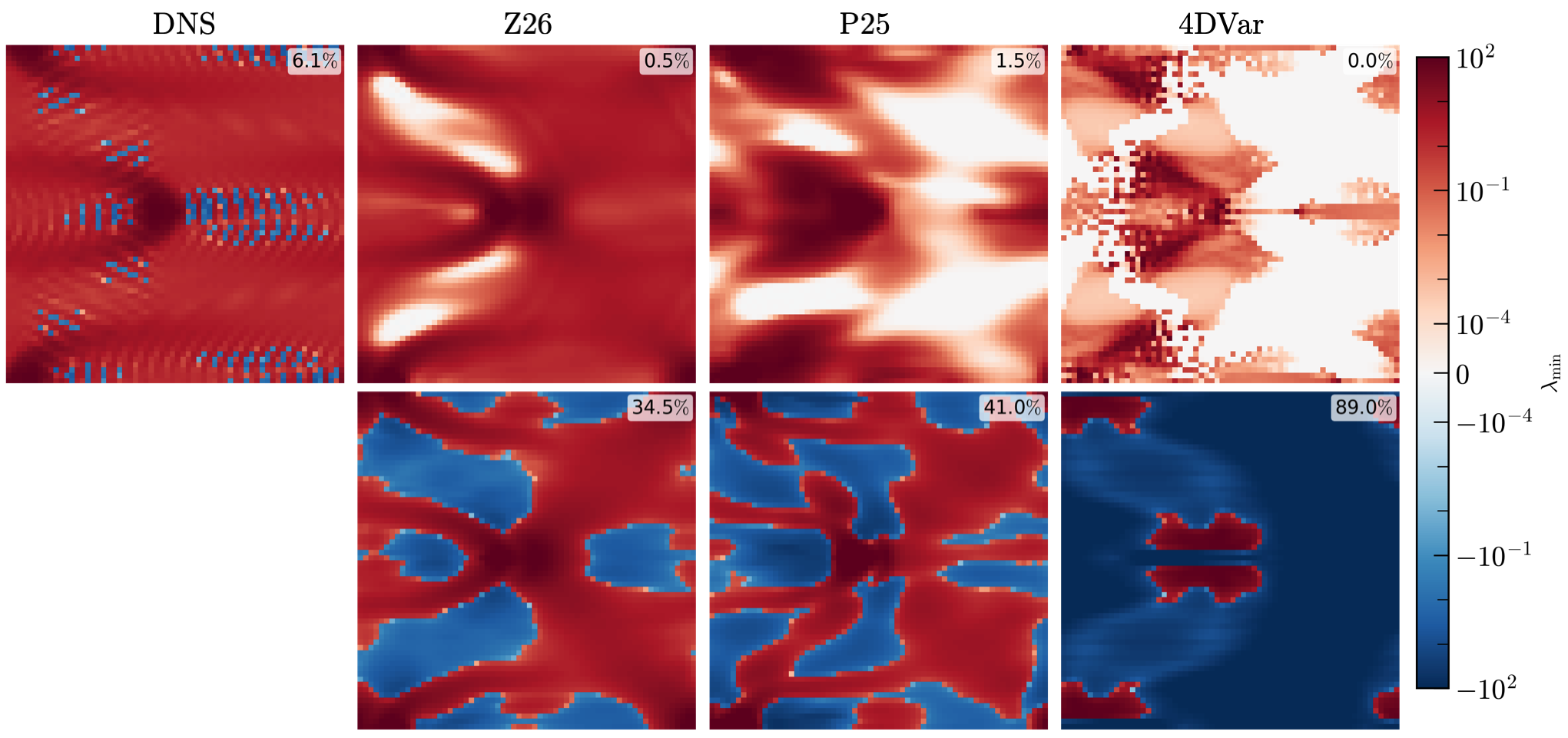}
    \caption{Minimum eigenvalue of the conformation tensor for the DNS and network predictions in the periodic-orbit case. Blue regions denote loss of positive definiteness (PD). The number in each subplot indicates the percentage of spatial grid points that violate PD. \JP{Top row shows the output of a network/4DVar algorithm with the positive definite preserving latent variables (equation \ref{eq:chol}), bottom row is the output of networks trained without this hard constraint. Note that the small numbers of points `violating' PD in the top row are due to minimum eigenvalues which are negative but on the order of machine precision.}}
    \label{fig:flowf_pd_po}
\end{figure}
\JP{
In figure 8 we show the importance of the Cholesky layer on maintaining physically realistic predictions with positive-definite conformation: we report the minimum eigenvalue of $\boldsymbol \alpha$ produced by the various state estimation methods with (top row) and without (bottom row) the hard constraint enforced. 
Note that the input DNS field itself has small patches where $\lambda_{min} < 0$; these areas can be removed via increase in spatial resolution and a drop in timestep and do not have discernible impact on the flow structures or statistics \citep[see recent discussion of the impact of loss of postivity in][]{Capocci2026}.
}
For the cases using the Cholesky parametrization, positive definiteness is guaranteed analytically by reconstruction, and the minimum eigenvalue field remains positive throughout almost the entire domain \JP{(there are small number of points where the minimum eigenvalue is negative, but is on the order of numerical roundoff)}. 
In contrast, without the Cholesky parametrization, a substantial fraction of the domain -- almost its entirety for 4DVar -- exhibits negative minimum eigenvalues, indicating violation of positive definiteness. 
Owing to the relative simplicity of the PO state, however, the forward-time integration remains stable despite these violations. 
At later times, positive definiteness is partially recovered as the oscillations of the initial prediction are smoothed by the forward mapping.

\subsection{Chaotic state}\label{sec:cs}
In the final case, we consider the elastic chaotic state (CS) observed by \citet{lewy2024revisiting} \JP{in a larger domain $6\pi \times 4\pi$}. 
In this state, multiple arrowhead structures collide and interact irregularly, producing an elasticity-driven chaotic flow with complex spatiotemporal dynamics. 
Owing to these features, reconstructing the conformation tensor is substantially more challenging and relies critically on maintaining the positive definiteness of the predicted tensor in order to prevent numerical failure of the forward-time solver.
For this case, the P25 and Z26 models without the Cholesky parametrization fail after several training iterations. 
In contrast, with the Cholesky parametrization, both P25 and Z26 remain stable, converge effectively, and produce physically reasonable conformation-tensor fields. 
For 4DVar, however, the optimization fails even when the positive definiteness of the initial state is enforced through the Cholesky parametrization. 
\JP{The breakdown of these predictions under time marching are associated with the high-wavenumber noise introduced into the initial state in the optimisation. 
This is a well-known feature of 4DVar and several methods have been proposed to alleviate the effect \citep[see e.g.][]{Wang2019}. 
}
\JP{Rather than attempt to remedy this problem here, }we compare only the performance of Z26 and P25 \JP{for the CS case}.

\JP{
Note that in this section we consider the reconstructions of the stress within the relatively short training dataset (which consists of a single, 400 time unit trajectory) -- i.e. treating the problem as a single-trajectory inverse problem.
However, we find that even with this limited data our Z26 model generalises reasonably well, which is something we explore further in the final subsection below.
}

\begin{figure}
      \centering    
    \includegraphics[width=0.7\linewidth, trim=0mm 0mm 0mm 0mm, clip]{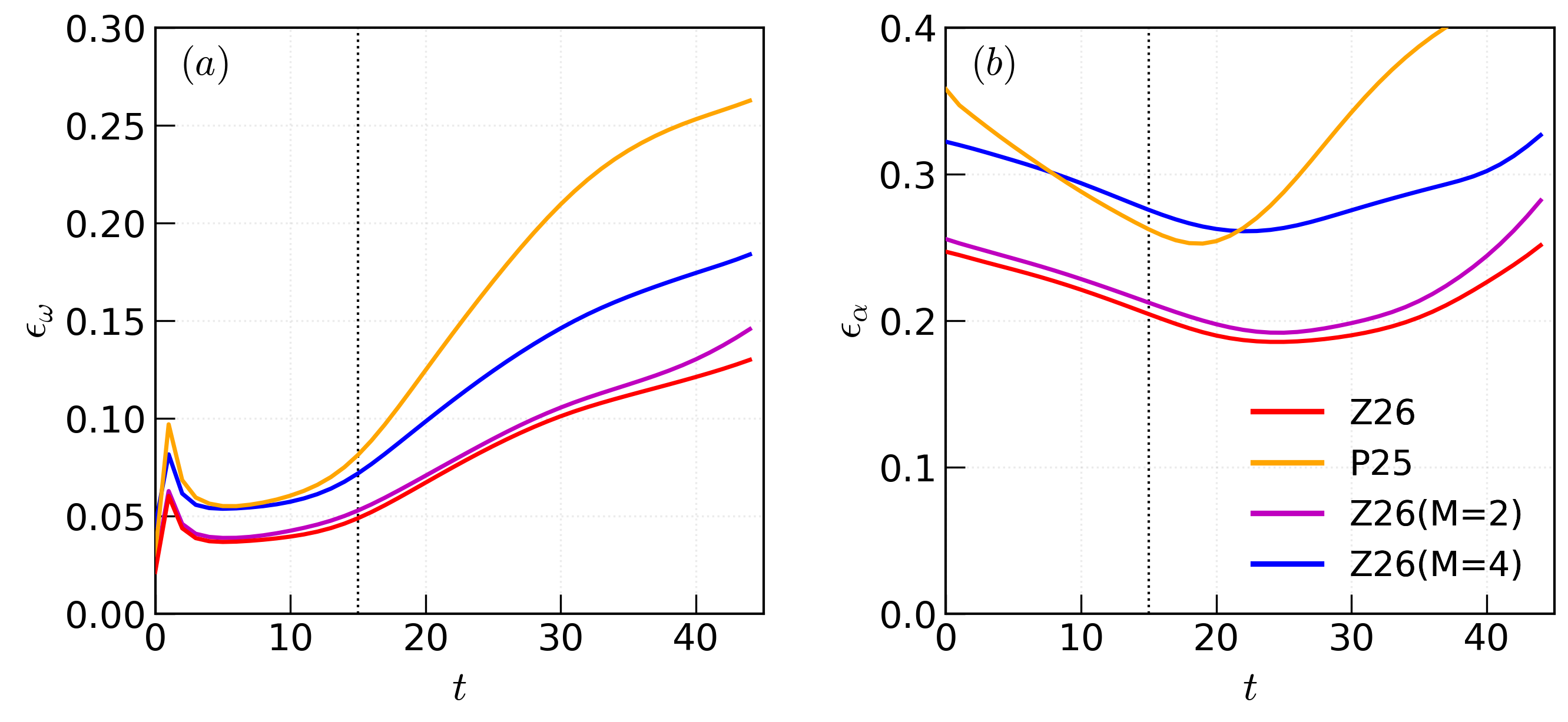}
    \caption{Temporal evolution of the relative reconstruction errors for the vorticity, $\epsilon_\omega$, and conformation tensor, $\epsilon_\alpha$, in the chaotic state (CS) case. The dotted line marks the assimilation time $T_e$.}
    \label{fig:error_cs}
\end{figure}
In figure \ref{fig:error_cs}, we \JP{report} the temporal evolution of the vorticity and conformation-field errors. 
Similar to the TW and PO cases, the vorticity reconstruction error, $\epsilon_\omega$, remains small, typically below $0.1$, whereas the conformation-tensor error, $\epsilon_\alpha$, are larger.
\JP{However, the performance is still comparable to the simpler dynamical states ($\epsilon_{\alpha} \sim 0.2$ for the best-performing Z26 network), while} the differences between cases are smaller than those observed for the TW and PO states, despite the greater complexity of the CS flow. 
This may be attributed to the more diverse and larger training dataset covered by the chaotic state.
\JP{On the other hand, unlike the simpler dynamics considered above, we generally observe a rise in all errors as time is marched beyond the assimilation/training time horizon, consistent with the chaotic nature of the attractor.}

\begin{figure}
      \centering    
    \includegraphics[width=0.7\linewidth, trim=0mm 0mm 0mm 0mm, clip]{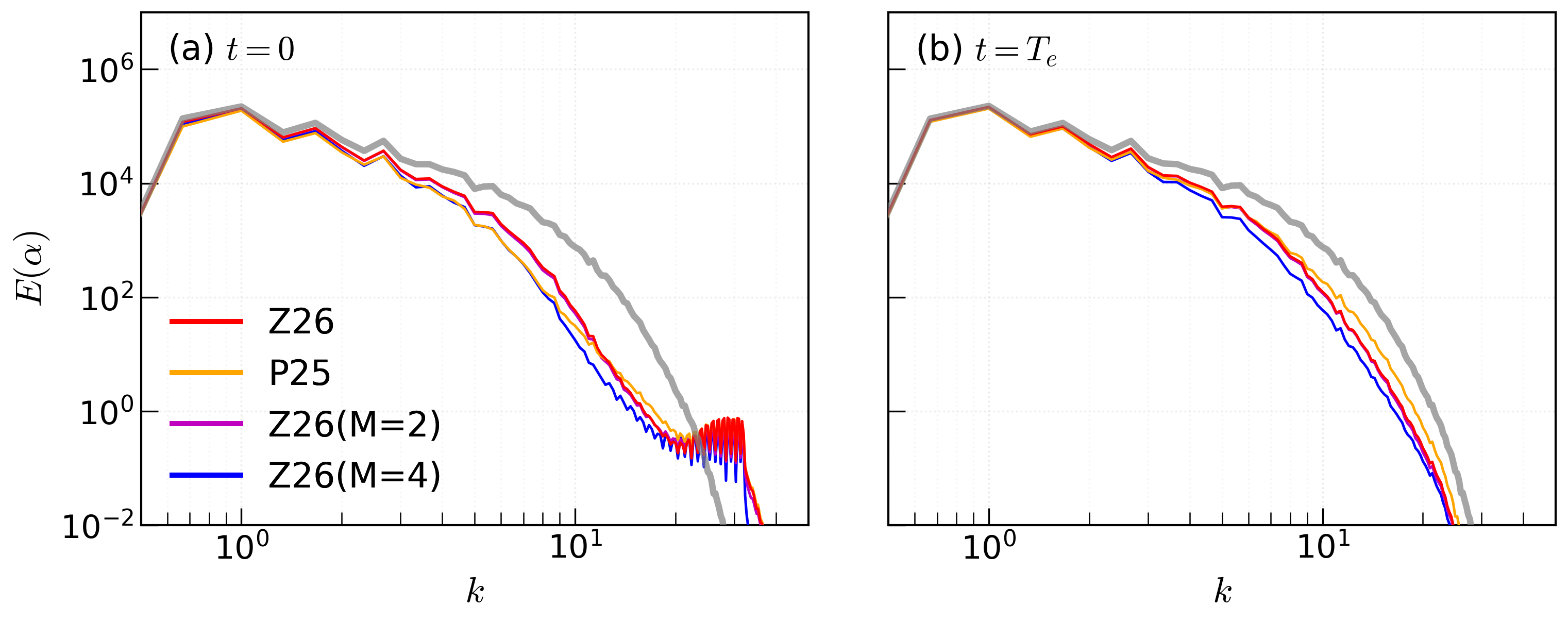}
    \caption{Spectral content of $\alpha_{xx}$ at (a) $t=0$ and (b) $t=T_e$ for \JP{an example} chaotic state. \JP{The DNS reference is shown by the tick grey line.}}
    \label{fig:spectra_cs}
\end{figure}
\JP{The reproduction of the conformation field by the neural networks under both Z26 and P25 algorithms is explored in figure \ref{fig:spectra_cs}, where we examine the total spectral content of $\alpha_{xx}^2$.}
All cases resolve the large-scale spectral content reasonably well up to approximately $k\sim 2$ (\JP{the scale associated with the large-scale elastic coherent structures}). 
\JP{The network trained via the Z26 loss (\ref{eq:loss:z26}) again produces the most faithful reconstruction.}
At higher wavenumbers, however, all predicted spectra are systematically lower than the DNS spectrum, reflecting the networks' underestimation of smaller-scale, or localised elastic fluctuations.
In the chaotic state, the flow dynamics is primarily governed by the large-scale arrowhead structures and the underestimation of small-scale fluctuations does not strongly affect the predicted trajectory \JP{at least at early times}. 
As the rollout proceeds, the reconstructed spectra gradually approach the DNS spectrum \JP{across the scales by} $t=T_e$.

\begin{figure}
      \centering    
    \includegraphics[width=\linewidth, trim=0mm 0mm 0mm 0mm, clip]{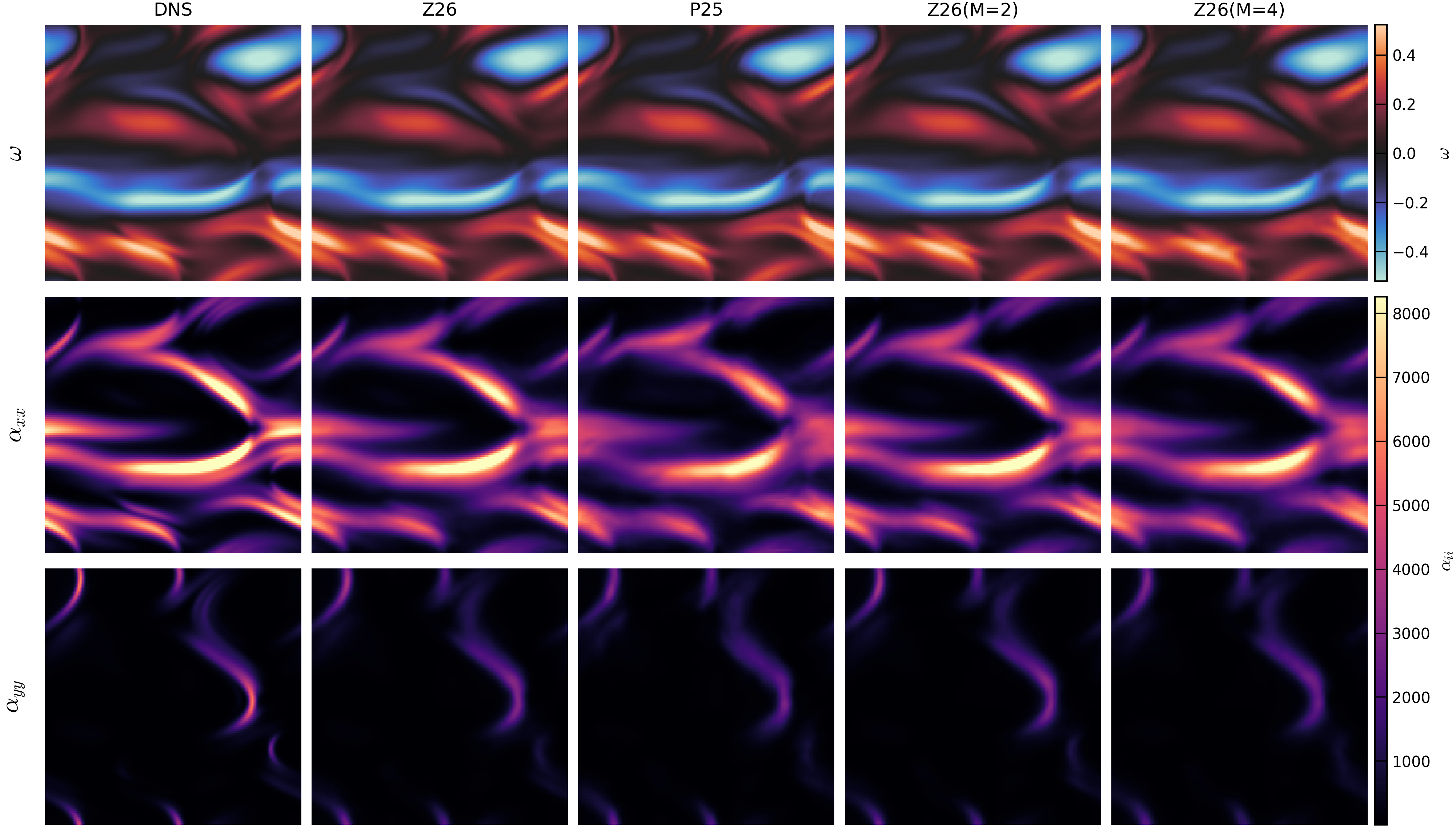}
    \caption{Comparison of network-predicted snapshots of individual flow fields with the corresponding DNS snapshots for the chaotic state. \JP{Rows show vorticity, streamwise-normal and vertical-normal conformation fields respectively. Columns are labeled by the corresponding state estimator.}}
    \label{fig:flowf_CS}
\end{figure}
\JP{Predicted vorticity and conformation fields in the CS regime are examined in figure \ref{fig:flowf_CS}. 
Unlike earlier examples in simpler dynamical regimes, all results reported in figure \ref{fig:flowf_CS} reproduce the qualitative features of the input, consistent with the lower relative errors and spectral content visualised above in figures \ref{fig:error_cs} and \ref{fig:spectra_cs}. 
Perhaps the most notable shortcoming is the under-prediction of the vertical conformation $\alpha_{yy}$, with the Z26 approach most closely matching the amplitude in the reference DNS. 
The strong performance of all cases, including the coarse grained Z26 results, can again be attributed to the richer training set.
}

\begin{figure}
      \centering    
    \includegraphics[width=0.9\linewidth, trim=0mm 0mm 0mm 0mm, clip]{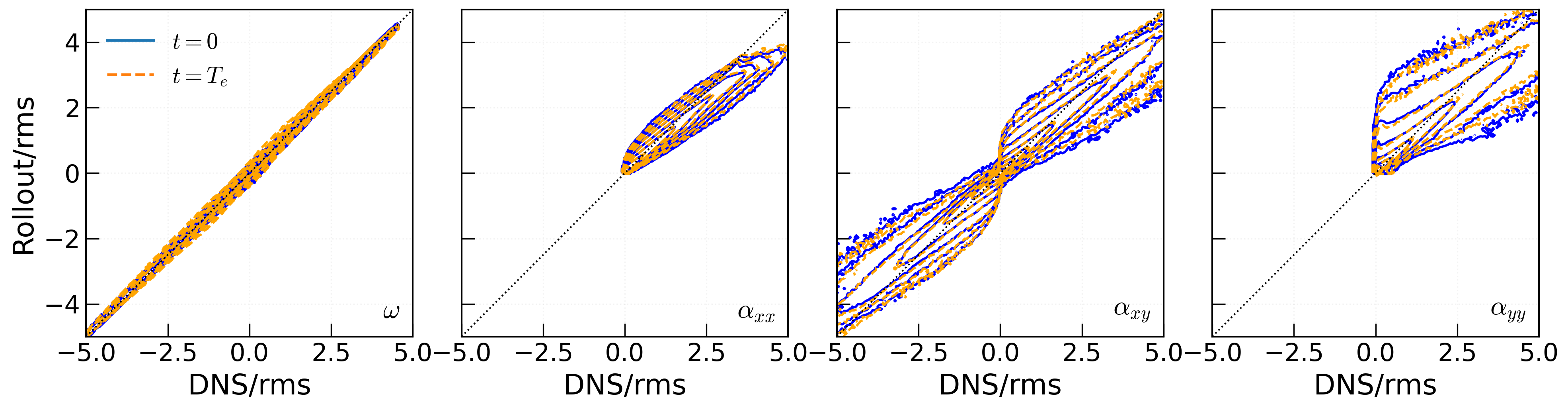}
    \caption{Joint PDFs between the DNS fields and predictions \JP{from the Z26 model are shown} for the predicted state at $t=0$ and \JP{after time marching to} $t=T_e$. All quantities are normalized by the root-mean-square value of the corresponding DNS field. The black dashed line marks the one-to-one line \JP{on which perfect predictions would lie}.}
    \label{fig:jpdf_CS}
\end{figure}
Consistent with the instantaneous-field comparisons, the joint PDFs between the DNS and Z26 rollout fields in figure \ref{fig:jpdf_CS} show that the model tends to underestimate \JP{smaller components of} the conformation tensor ($\alpha_{xy}$ and $\alpha_{yy}$), particularly \JP{when the true state is larger in amplitude}.
\JP{The same effect is seen to a less extent with $\alpha_{xx}$, but for all examples the}
joint-PDF contours of the conformation components $\alpha_{xx}$, $\alpha_{xy}$, and $\alpha_{yy}$ bend below the one-to-one line, indicating underprediction of the extreme values. 
By contrast, $\alpha_{yy}$ is slightly overestimated in the low-magnitude region.

\subsection{Model extrapolation}\label{sec:box}
In the previous section, the reconstruction model was trained and tested only in a minimal domain unit of size $L_x\times L_y=6\pi\times 4\pi$. 
In practical experimental settings, the domain size may be much larger \JP{and} network training \JP{on the full problem may be prohibitively expensive}. 
On the other hand, previous studies have shown that arrowhead structures exhibit characteristic and reproducible conformation and velocity configurations~\citep{zhuJFM2026on} \JP{with well defined length scales.}
\JP{This indicates that it may be possible to make conformation predictions in larger domains using our ``minimal'' chaotic network described above.}
To examine this possibility, in this section we test the transferability of Z26 model trained above to larger domains of size $L_x\times L_y=8\pi\times 8\pi$ and $16\pi\times 16\pi$ (chaos is sustained in both boxes). 
\JP{We do this without any retraining of the model, which is possible due to its purely convolutional nature and the local filters which it applies.}
\JP{We also note that the model we are examining here was trained on a single, relatively short time series (total 400 time units); larger-scale training will likely produce a more robust model, something we discuss further in our conclusions.}

\begin{figure}
      \centering    
    \includegraphics[width=0.4\linewidth, trim=0mm 0mm 0mm 0mm, clip]{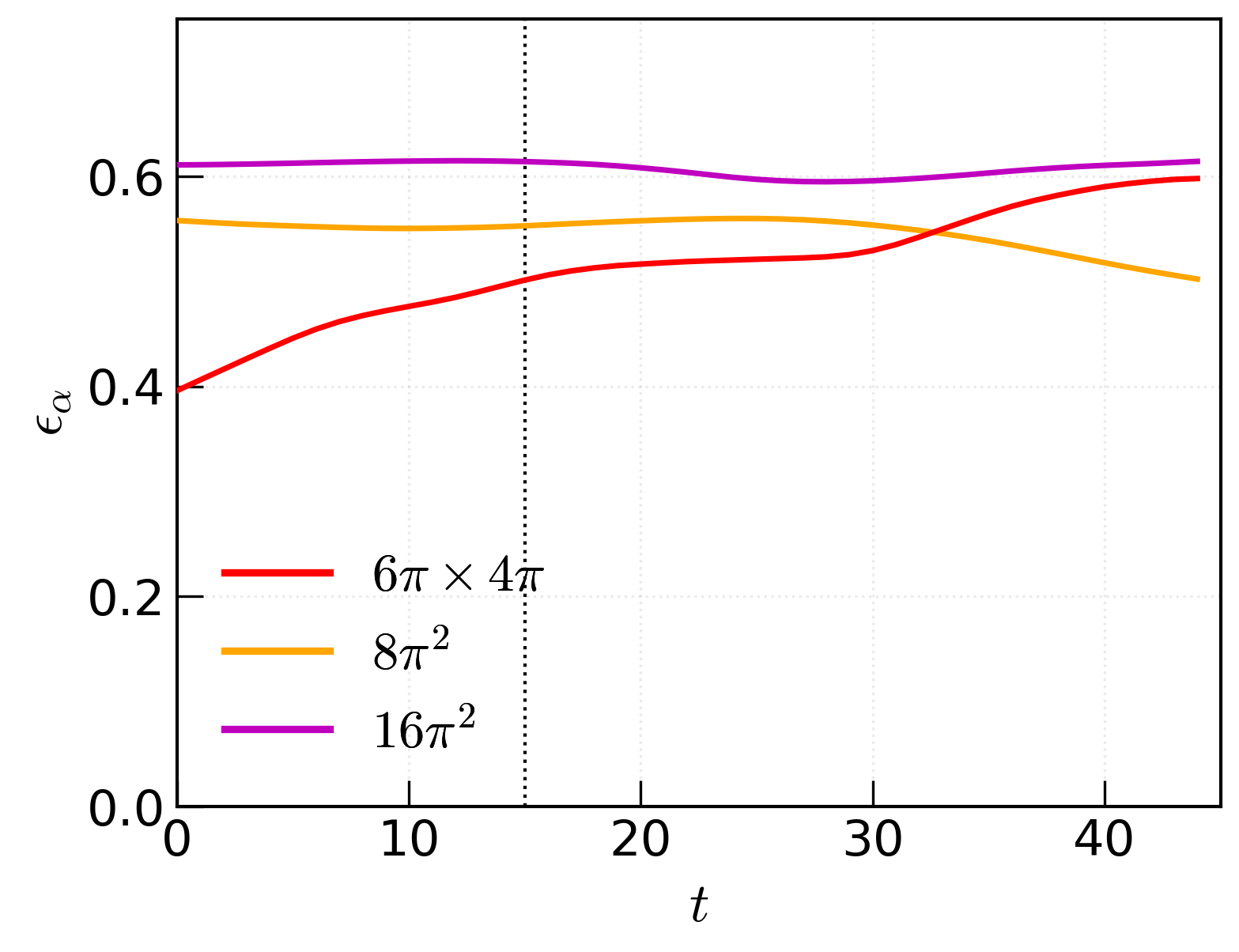}
    \caption{Temporal evolution of the relative reconstruction error for the conformation tensor, $\epsilon_\alpha$, in the model extrapolation study. The Z26 network trained on the short 400 time unit trajectory is applied to much a variety of domain sizes. The dotted line marks the assimilation time $T_e$ (time-marching horizon time used in training of the original model).
    }
    \label{fig:error_box}
\end{figure}

In figure \ref{fig:error_box}, we show the temporal evolution of the conformation-tensor errors for three domain sizes: $6\pi\times 4\pi$, using a trajectory independent of the training set considered in the previous section, and the larger domains $(8\pi)^2$ and $(16\pi)^2$. 
Comparing the different domain sizes, the $6\pi\times 4\pi$ case clearly gives the best performance, although its initial conformation error, $\epsilon_\alpha \sim 0.4$ at $t=0$, remains larger than that of the corresponding Z26 case in \S\ref{sec:cs}, where $\epsilon_\alpha\sim 0.25$. 
\JP{This is expected due to the very small amount of training data -- a single 400 time unit trajectory -- which was motivated by both (i) consideration first of a single-trajectory inverse problem and (ii) potential limitations in the amount of data that may be available in an experiment.}
For the larger domains, the errors are greater than those in the $6\pi\times 4\pi$ case, but are surprisingly close to one another. 
This similarity may reflect the fact that the arrowhead structure acts as a basic building block of the chaotic state\JP{, and that as a result the model may only be weakly dependent on the overall box size}.

\begin{figure}
      \centering    
    \includegraphics[width=0.97\linewidth, trim=0mm 0mm 0mm 0mm, clip]{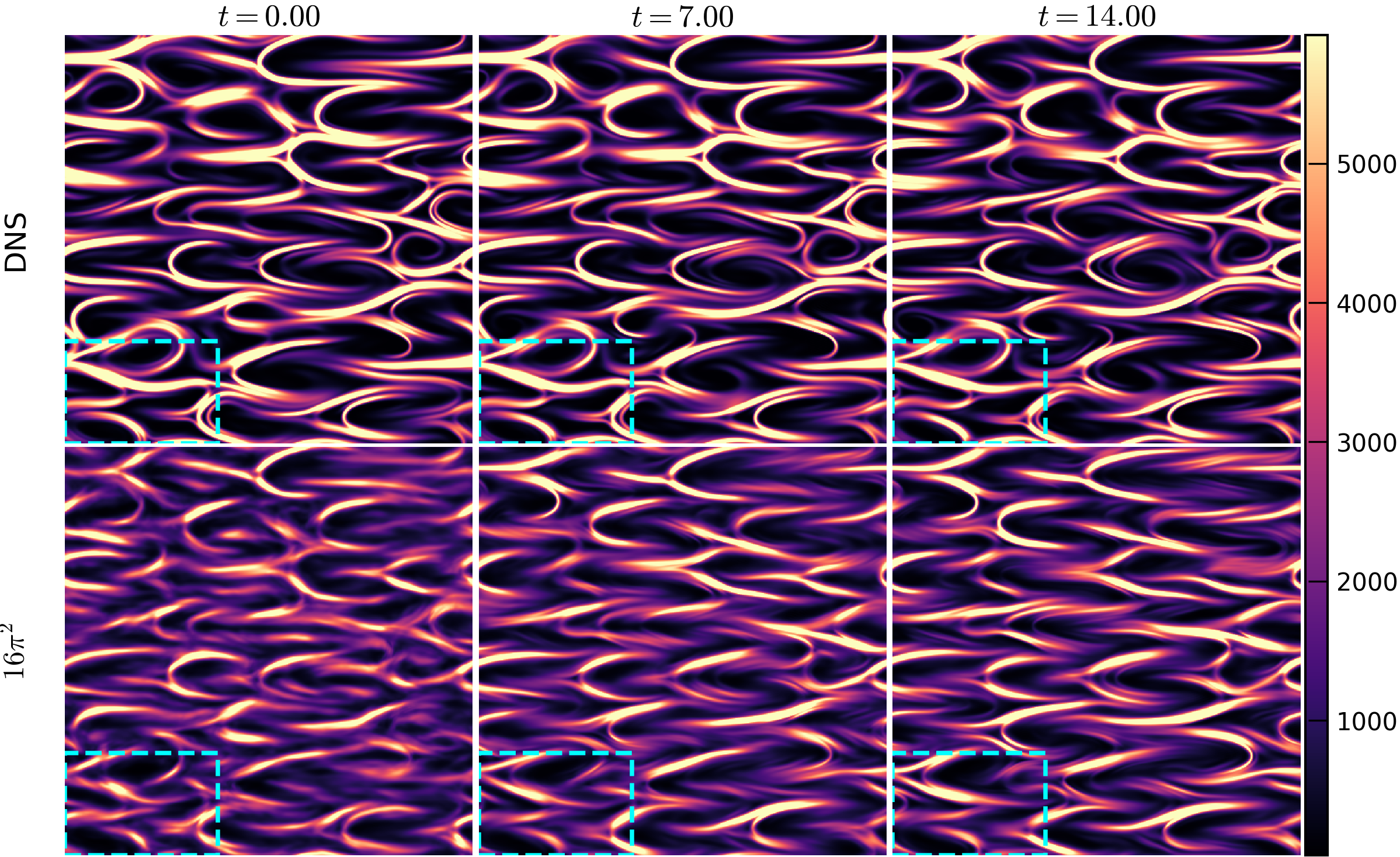}
    \caption{Temporal evolution of the trace of the conformation tensor, $\mathrm{tr}\,\bm{\alpha}$, along the rollout trajectory for the travelling-wave case at $t=0$, $7$, and $14$. Blue-dash box indicates the domain size of the training dataset.}
    \label{fig:flowf_box}
\end{figure}
\JP{A full field comparison of $\text{tr}\,\boldsymbol \alpha$ to DNS is shown in figure \ref{fig:flowf_box}, where we visualise the initial network reconstruction and the subsequent evolution in time.}
\JP{Despite the larger relative errors observed in figure \ref{fig:error_box}, the reproduction of the conformation field is relatively robust, with clear reproduction of highly stretched regions and the signature shapes of the elastic coherent structures. 
The stretch amplitude is typically weaker than the full DNS, and the sheets more diffuse, but the sharpness of the structures improves as the prediction is marched forward in time. 
This bodes well for reproduction of new Kolmogorov geometries, particularly if additional trajectory data (still in the form of vortical time series) are available for training of the minimal model. 
}

\section{Conclusions}\label{sec:concl}
\JP{
In this paper we have presented a new method to estimate the polymer conformation tensor given only measurements of vorticity. 
Our approach uses a convolutional neural network to output a strictly positive definite conformation field (as a hard constraint) given a five-snapshot vorticity series (input as a single, 5-channel image), combined with a training algorithm inspired by variational assimilation methods. 
This online learning approach calls a differentiable solver in the evaluation of the loss, which propagates the network’s predictions forward in time for comparison to the stored data. 
In addition to a measurement term, we include also a `self-consistency’ component in the loss \citep{Zhu_Page2026} which compares network predictions on later-time measurements to the solver-predicted state. 
We examined the network for three different flow regimes — a traveling wave, periodic orbit and a fully chaotic state. 
The network was trained on a small amount of data — as a single-trajectory inverse problem — and was found to be very effective in estimating the conformation, both in terms of structures and the absolute values of the stretch. 
We observed substantial improvements over the standard 4DVar algorithm and another simpler online learning approach. 
}

\JP{
Despite the small training dataset, the network trained on chaotic data was found to generalise reasonably well to new geometries, reproducing the sheet-like coherent structures in the polymer conformation. 
This is useful from the point of view of examining experimental data in varying aspect ratios and also highlights that the fundamental length scale in the problem is tied to the arrowhead coherent structure rather than the box size. 
A key next step is to widen the training dataset to include multiple independent trajectories which should aid in this generalisation. 
Another key unknown in an experiment is the polymer model and its underlying parameters. 
Our method has an advantage in that no offline data generation occurs, and in principle the model parameters can also be `learned’ as network outputs. 
Consideration of this kind of approach, with similar ideas seen in rheometric-flow results in \citet{Alp2025}, is something we are actively exploring. 
}

\bibliography{FPC,General,ML,VE}

\end{document}